\documentclass[ajl]{emulateapj}

\usepackage{natbib}
\usepackage{graphicx}
\usepackage{subfigure}
\usepackage{amsmath,amssymb}
\usepackage{color}
\usepackage{rotating}
\usepackage{hyperref}				
\hypersetup{colorlinks,%
	linkcolor=blue,%
	citecolor=blue}
\slugcomment{Accepted for publication in ApJ Letters}

\begin{document}
\title{Two Types of Long-duration Quasi-static Evolution of Solar Filaments}

\author{C. Xing$^{1,2}$, H. C. Li$^{1,2}$, B. Jiang$^{1,2}$, X. Cheng$^{1,2}$, M. D. Ding$^{1,2}$}
\affil{$^1$School of Astronomy and Space Science, Nanjing University, Nanjing, 210046, China}\email{xincheng@nju.edu.cn}

\affil{$^2$Key Laboratory of Modern Astronomy and Astrophysics (Nanjing University), Ministry of Education, Nanjing 210093, China}

\begin{abstract}
In this Letter, we investigate the long-duration quasi-static evolution of 12 pre-eruptive filaments (4 active region and 8 quiescent filaments), mainly focusing on the evolution of the filament height in three dimension (3D) and the decay index of the background magnetic field. The filament height in 3D is derived through two-perspective observations of \textit{Solar Dynamics Observatory} and \textit{Solar TErrestrial RElations Observatory}. The coronal magnetic field is reconstructed using the potential field source surface model. A new finding is that the filaments we studied show two types of long-duration evolution: one type is comprised of a long-duration static phase and a short slow rise phase with a duration of less than 12 hours and a speed of 0.1--0.7 km s$^{-1}$, while the other one only presents a slow rise phase but with an extremely long duration of more than 60 hours and a smaller speed of 0.01--0.2 km s$^{-1}$. At the moment approaching the eruption, the decay index of the background magnetic field at the filament height is similar for both active region and quiescent filaments. The average value and upper limit are $\sim$0.9 and $\sim$1.4, close to the critical index of torus instability. Moreover, the filament height and background magnetic field strength are also found to be linearly and exponentially related with the filament length, respectively.
\end{abstract}
\clearpage

\section{Introduction}
Solar filaments are cool and dense plasma suspended in the hot and tenuous corona, also called as prominences when observed on the solar limb \citep{2010SSRv..151..333M}. Based on distinct magnetic environments, filaments are usually grouped into active region (AR) filaments and quiescent (QS) filaments \citep{McCauley2015Prominence}. Many studies have been devoted to the physical properties of the two types of filaments including their length, height, and magnetic field strength. QS filaments are typically less than 30 Mm high and 200 Mm long \citep{2015ASSL..415...31E}. The magnetic field strength is about 3-15 G \citep{2010SSRv..151..333M}. By contrast, the magnetic field strength for AR filaments is usually hundreds of G \citep{1974SoPh...39..107T,2009A&A...501.1113K}. The height normally does not exceed 10 Mm \citep{2000AstL...26..322F} and the length is typically $\sim$30 Mm \citep{2000eaa..bookE2282M}.

The eruptions of filaments are usually associated with coronal mass ejections (CMEs) and flares as disclosed by many statistical studies \citep{Subramanian2001Source,Chandra2010How}. The eruption process generally has two phases: a slow rise phase with a speed of about several km s$^{-1}$ \citep{2004ApJ...602.1024S,2007SoPh..246...89I,2011A&A...533L...1R} and a rapid acceleration phase with an average acceleration of more than 1 km s$^{-2}$ \citep{Song2015Acceleration}. The duration of the slow rise phase varies from event to event. It lasts for several hours for QS filaments \citep{2004ApJ...602.1024S,McCauley2015Prominence}, but only tens of minutes for AR filaments \citep{2005ApJ...630.1148S,2006A&A...458..965C,McCauley2015Prominence}. Some events with a longer slow rise phase ($\sim$23 h) have also been found \citep[e.g.,][]{McCauley2015Prominence}.

\begin{table*}
\renewcommand\arraystretch{1.5}
\caption{Parameters of filaments at the near-eruption stage}
\begin{tabular}{ccccccccccc}
\hline
\hline
No. & Type & Start time & Final time & Lat ($^{\circ}$) & Lon ($^{\circ}$) & $L$ (Mm) & $H$ (Mm) & $n$ & $B_h$ (Gs) & $v$ (m s$^{-1}$) \\
\hline
F1&	QS&	2010/05/16 08:35&	2010/05/21 12:10&	8$\sim$32&	  -30$\sim$-9&	361&	54$\pm$4&	  1.4$\pm$0.1&	    6.8$\pm$0.8&	656$\pm$76 \\
F2&	QS&	2010/06/14 06:10&	2010/06/20 00:00&	18$\sim$25&	  0$\sim$18&	207&	36$\pm$3&	  0.79$\pm$0.07&    3.9$\pm$0.3&	340$\pm$61 \\
F3&	QS&	2010/07/25 23:50&	2010/08/01 10:50&	21$\sim$34&	  -2$\sim$12&	218&	44$\pm$10&	  0.61$\pm$0.14&    2.5$\pm$0.3&	199$\pm$119 \\
F4&	QS&	2010/08/06 10:50&	2010/08/14 07:00&	20$\sim$34&	  26$\sim$45&	251&	33$\pm$1&	  0.88$\pm$0.10&    7.8$\pm$1.0&	142$\pm$15 \\
F5&	AR&	2010/09/01 00:00&	2010/09/07 11:00&	19$\sim$21&	  8$\sim$12&	51&	    15$\pm$2&	  0.56$\pm$0.05&    20$\pm$1&	    205$\pm$54 \\
F6&	QS&	2010/12/18 00:00&	2010/12/20 23:24&	23$\sim$31&	  -54$\sim$-20&	360&	53$\pm$6&	  0.97$\pm$0.08&    1.3$\pm$0.1&	389$\pm$159 \\
F7&	AR&	2011/02/12 14:15&	2011/02/15 22:46&	20$\sim$32&	  -32$\sim$-24&	164&	34$\pm$1&     0.78$\pm$0.01&    14$\pm$0.3&	    658$\pm$34 \\
\hline
F8&	QS&	2010/10/10 00:00&	2010/10/18 10:20&	-31$\sim$-20& 25$\sim$50&	282&	31$\pm$2&	  0.56$\pm$0.04&    14$\pm$1&	    11$\pm$2  \\
F9&	AR&	2010/12/01 05:20&	2010/12/01 13:20&	12$\sim$15&	  -29$\sim$-28&	37&	    8.6$\pm$0.2&  0.74$\pm$0.02&    52$\pm$1&	    136$\pm$16 \\	
F10& QS& 2010/12/03 12:00&	2010/12/06 14:10&	-41$\sim$-12& -54$\sim$-6&	589&	56$\pm$3&	  1.2$\pm$0.04&	    8.0$\pm$0.5&	97$\pm$3  \\	
F11& QS& 2010/12/30 00:00&	2011/01/09 16:00&	-41$\sim$-22& 20$\sim$73&	559&	44$\pm$2&	  0.85$\pm$0.04&    7.4$\pm$0.4&	73$\pm$6  \\	
F12& AR& 2011/06/05 10:50&	2011/06/07 04:50&	-24$\sim$-19& 50$\sim$54&	67&	    16$\pm$0.2&   1.1$\pm$0.1&	    100$\pm$19&     52$\pm$1  \\
\hline
\vspace{0.005\textwidth}
\end{tabular}
Note: Parameters $L$ and $H$ are the filament length and height, respectively. $B_h$ and n are the strength and decay index of the horizontal component of the background magnetic field at the filament height, respectively. The parameter $v$ refers to the average speed in the slow rise phase for two groups of filaments.
\end{table*}

Prior to the eruption, filaments are in quasi-static equilibrium that generally lasts for hours to days \citep{1985SoPh..100..415H,Zuccarello2014Observational}. The equilibrium can be achieved when the upward force from the filament structure is balanced by the magnetic tension of the background field \citep{2006PhRvL..96y5002K,2011ApJ...732...87C}. In the context of the magnetic structure of filaments being a magnetic flux rope \citep[MFR;][]{1995ApJ...443..818L,2017ScChE..60.1383C}, the onset of the filament eruptions is possibly attributed to some MHD instabilities such as torus and/or kink instabilities. The torus instability takes place when the background field over the filament-MFR system declines fast enough \citep[e.g.,][]{2006PhRvL..96y5002K,2010ApJ...718..433O,Demoulin2010Criteria}. The kink instability occurs when the twist number of the MFR exceeds a threshold value of 1.25--1.75 turn \citep[e.g.,][]{1981GApFD..17..297H,Fan2003,Torok2004}. In addition, the eruption of the MFR-filament system can also be initiated by the tether-cutting \citep{2001ApJ...552..833M} and breakout \citep{1999ApJ...510..485A} reconnection, as which are able to increase the upward magnetic pressure and decrease the downward magnetic tension, respectively.

In this Letter, we investigate the long-duration ($\sim$3 days) evolution of 12 pre-eruptive filaments (4 AR and 8 QS filaments), primarily focusing on the evolution of their height in 3D and the decay index of the background magnetic field. In Section 2, we present observational data and methods. The results are shown in Section 3, which is followed by a summary and discussions in Section 4.

\begin{figure*}
\centering
\plotone{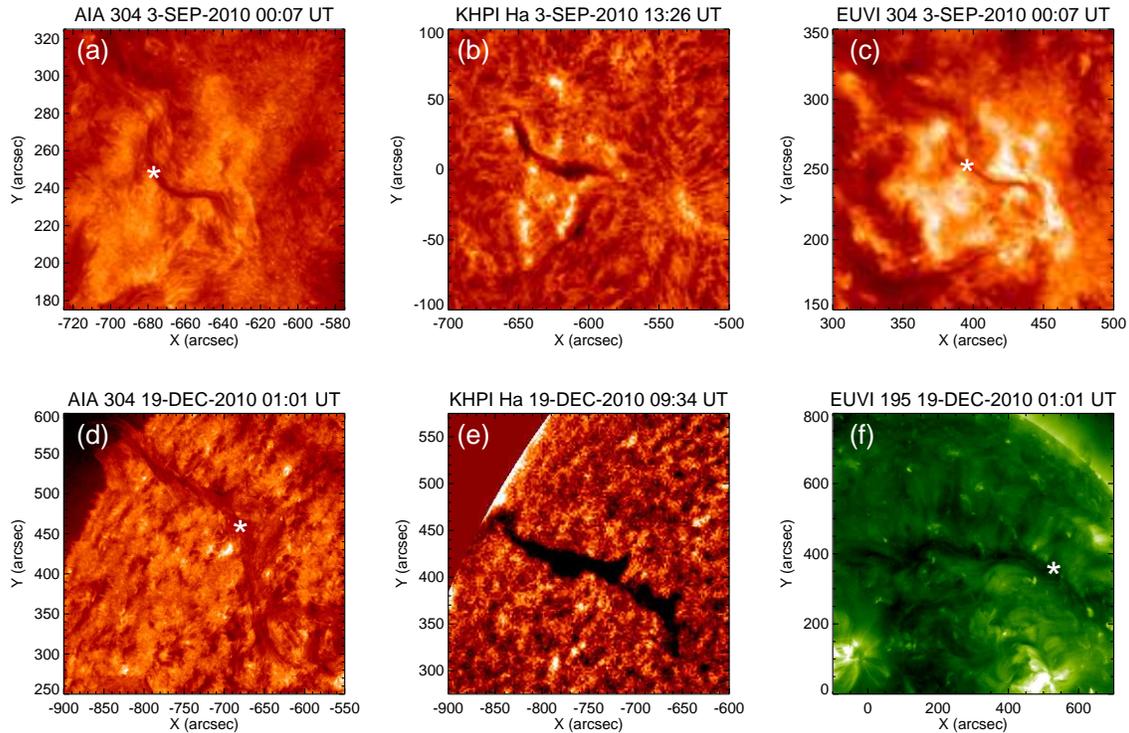}
\caption{\textit{SDO}/AIA and \textit{STEREO}/EUVI images at the 304 (or 195) {\AA} passband and H$\alpha$ images showing snapshots of the F5 and F6 filaments from two different perspectives. The stars in panels a and c (d and f) indicate the same feature in the F5 (F6) filament.}
\end{figure*}

\section{Observational Data and Methods}
We select eruptive filaments using the H$\alpha$ data provided by the full-disk H$\alpha$ patrol telescope at Big Bear Solar Observatory\footnote{http://www.bbso.njit.edu/NJIT\_Ground-Based\_Solar\_Observatories.pdf}. The time range is from May 2010 to June 2011, during which the separation angle between \textit{Solar Dynamics Observatory} (\textit{SDO}; \citealt{2012SoPh..275....3P}) and one satellite of \textit{Solar TErrestrial RElations Observatory} (\textit{STEREO}; \citealt{2008SSRv..136...67H}) (\textit{STEREO-A} or \textit{STEREO-B}, depending on the location of filaments) is less than 90 degrees. It allows us to determine the 3D parameters of filaments. The start time of filaments refers to the moment when they are first observed by SDO. The end time is the near-eruption time of the filaments, which means that the filaments will erupt within a time window of less than one hour. The first six columns of Table 1 display the basic information of each event including its type, start time, and end time.

\begin{figure*}
\centering
\plotone{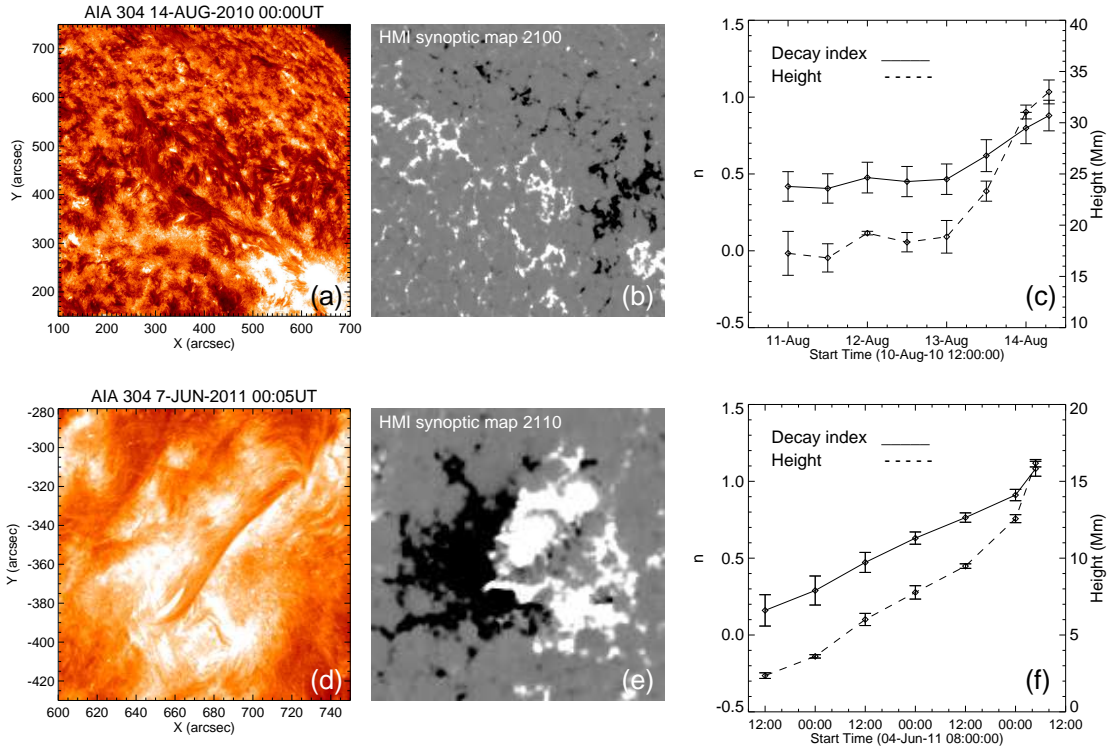}
\caption{(a) AIA 304 {\AA} images showing the F4 filament on 2010 August 14. (b) HMI synoptic magnetogram displaying the magnetic field distribution of the F4 source region. (c) Temporal evolution of the height in 3D and the corresponding decay index of the background magnetic field for the F4 filament. (d)--(f) The same as panels (a)--(c) but for the F12 filament.}
\end{figure*}

Our sample consists of 4 AR filaments and 8 QS ones, which are denoted by F1-F12. In the following, we take F5 and F6 as examples to interpret our procedure of reconstructing the parameters of filaments in 3D. F5 is an AR filament that was well observed from 2010 September 1 to 7. F6 is a QS one observed from 2010 December 18 to 20. In order to determine the locations of filaments in 3D, we use the simultaneous observations of the Atmospheric Imaging Assembly (AIA; \citealt{2012SoPh..275...17L}) on board \textit{SDO}, which images the solar corona with a pixel size of 0.6" and a cadence of 12 s, and of the Extreme UltraViolet Imager (EUVI; \citealt{2004SPIE.5171..111W}) on board \textit{STEREO}, which has a pixel size of 1.6" and a cadence of 10 min. The passband selected is 304 {\AA}, at which the filaments can be easily discerned (Figure 1). The code \texttt{SCC\_MEASURE} in the SolarSoftWare (SSW) developed by W. Thompson is used to measure the 3D positions of the filaments. For each perspective, a straight line connecting the satellite and one specific feature of the filaments can be drawn. The cross point of the two lines is regarded as the 3D position of the filament feature. By repeating the same procedure, the 3D positions of a sequence of features from one end to the other end of the filaments are obtained. The filament length is calculated as the sum of the distances between any two adjacent features. The filament height is considered to be the average height of the middle part of the filament, which has the highest height for most of events. The minimal and maximal values of longitudes and latitudes, the heights, and the lengths for all filaments at the near-eruption time are listed in Table 1. 

\begin{figure*}
\centering
\plotone{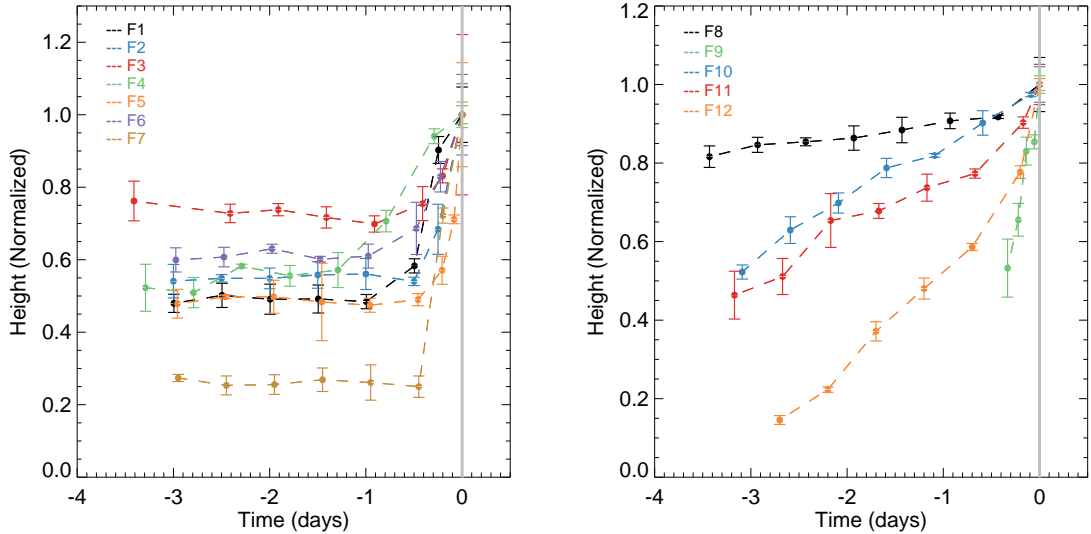}
\caption{Temporal evolution of the height for the group 1 (left) and group 2 (right) filaments. The height for each event is normalised by that at the near-eruption stage. The vertical lines in both panels indicate the near-eruption time.}
\end{figure*}

We also calculate the decay index $n$ of the background field, which is believed to be a key factor to determine the initiation of the filaments \citep{2005ApJ...630L..97T,2007ApJ...668.1232F,Liu2008Magnetic}, by the following formula:
\begin{equation}
\centering
n=-\frac{\rm d(ln\it B_h)}{\rm d(ln\it h)}
\end{equation}
where $B_h$ represents the horizontal component of the background magnetic field and $h$ denotes the height. The 3D coronal magnetic field is reconstructed by the potential field source surface model \citep[PFSS;][]{1969SoPh....6..442S,1969SoPh....9..131A,2017ScChD..60.1408G}. We use the code in the PFSS package in SSW, which assumes that the magnetic field is potential between the photosphere and spherical surface and that the magnetic field on the spherical surface is radial. The spherical surface is set as 2.5 solar radii in this work. The Carrington synoptic maps constructed from HMI line-of-sight magnetograms observed by Helioseismic and Magnetic Imager (HMI; \citealt{2012SoPh..275..207S}) on board \textit{SDO} are used as the bottom boundary. The original synoptic maps are then resampled with a resolution of 6 Mm, the grid in the vertical direction is set to increase from 0.35 to 3.5 Mm with the filament height. To derive the decay index $n$ at the filament height, we first make an average of the background magnetic field $B_h$ over the main polarity inversion line (PIL) in the filament source regions. The errors of $B_h$ and $n$ are mainly from the uncertainties in height, which are regarded as the standard deviations of a number of measurements. The horizontal magnetic field strengths and decay indices at the filament heights are also listed in Table 1. Note that, all filaments we selected are far away from the solar limb, thus the calculation of the background magnetic field is not significantly influenced by the discontinuity of the magnetic field on the solar limb in the synoptic map.

\section{Results}
\subsection{Two Types of Quasi-static Evolution}
We take a QS filament (F4) and an AR filament (F12) as two examples to interpret how the filament heights and the corresponding decay indices evolve in the quasi-static evolution phase prior to their eruptions (Figure 2). One can see that the F4 filament always stays at a similar height in 3D (17-19 Mm) with a similar decay index ($\sim$0.43) from 2010 August 11 to 13. After that, the filament rapidly ascends to a height of 33 Mm at the moment approaching the eruption. The decay index increases from 0.43 to 0.88 correspondingly (Figure 2c). For the F12 filament, however, its height has been increasing linearly from 2 Mm on 2011 June 4 to 16 Mm on 2011 June 7. The corresponding decay index increases from 0.16 to 1.08 in that time period (Figure 2f).

In Figure 3, we show the evolution of the heights for 12 filament events in the period of 3 days before the eruption. One can clearly see that the evolutions of 12 filaments can be divided into two types: one experiences a long-duration static phase and then a short-duration slow rise phase (group 1; left panel); the other only presents a slow rise phase but with an extremely long duration (group 2; right panel). For the group 1 filaments, during the long-duration static phase, the filaments keep in a stable state without significant change in height. The filaments do not start to ascend until entering the short-duration slow rise phase. The duration of the slow rise phases is shorter than 12 h except for the F4 event ($\sim$30 h). By contrast, for the group 2 filaments, there is no static phase in the three days prior to the eruption. Those filaments only display a slow rise phase but with an extremely long duration. Except for the F9 event, the duration of the filaments is at least 60 h, much longer than that of the slow rise phase for the group 1 filaments. 

We also calculate the average speed in the slow rise phase for the two groups of filaments. The average speed is obtained by a linear fitting to the height-time data in the slow rise phase, and the error of the speed comes from the error of the height. The speed in the short-duration slow rise phase for the group 1 filaments ranges from 0.1 to 0.7 km s$^{-1}$. For the group 2 filaments, the average speed is only 0.01--0.2 km s$^{-1}$, systematically smaller than that for the group 1 filaments. Such a distinction indicates that there may exist two different physical processes to control the quasi-static evolution of pre-eruptive filaments.

It is worth addressing that the F9 filament is a special event, which first appears at 05:20 UT on 2010 December 1 and then erupts shortly after 13:20 UT on the same day. In the time period of 8 h, its height increases continuously and linearly. Because of the absence of a static phase, we classify it as a group 2 filament. However, compared with other group 2 filaments, this event has a much shorter slow rise phase, thus resulting in a larger speed that is even comparable to the average speed for the group 1 filaments.

\begin{figure*}
\centering
\plotone{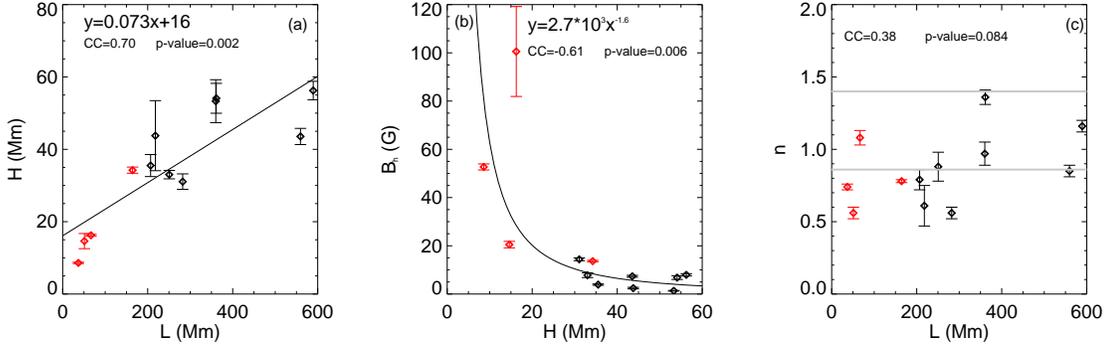}
\caption{(a) Scatter plot of the filament height versus the length. The oblique line displays the linear fitting to the data points. The red and black points refer to the AR and QS filaments, respectively. (b) Scatter plot of the background magnetic field strength versus the height. The curve shows the exponential fitting. (c) Scatter plot of the decay index at the filament height versus the length. The grey lines show the upper limit and the average value.}
\end{figure*}

\subsection{Properties at the Near-eruption Stage}
We further study the relationship between the parameters of filaments prior to the eruption. Figures 4a and 4b show the scatter plots of the filament length and the background magnetic field strength versus height, respectively. One can see that the three parameters have a broad distribution. The AR filaments have a length of 37--165 Mm, much smaller than the length of the QS ones (207--590 Mm). The height of the AR filaments is in the range of 8--35 Mm, also lower than that of the QS ones (30--57 Mm). The background magnetic field also appears differently for the distinct types of filaments. Overall, the background magnetic field for the AR filaments (10-100 G) is stronger than that of the QS ones (1-14 G). Moreover, the filament height is found to vary linearly with the length with a fitting function of $H=0.073L+16$, where $L$ and $H$ are expressed in Mm. Their correlation coefficient (Kendall's tau) is about 0.70 with a p-value of 0.002 ($<$0.05). The background magnetic field strength at the filament height is found to vary exponentially with the filament height. The corresponding fitting function is $B_h=2.7\times10^{3}H^{-1.6}$, where $H$ is in Mm and $B_h$ is in G. The correlation coefficient between them is about --0.61 with the p-value of 0.006 ($<$0.05). These results imply that the AR and QS filaments may have similar magnetic environments and the apparent discrepancies may be mainly from the quantitative difference in the magnetic field strength.

Figure 4c shows the scatter plot of the decay index versus the filament length. One can find that the decay index at the near-eruption stage is independent of the filament length. The correlation coefficient between the decay index and filament length is  0.38 with the p-value of 0.084 ($>$0.05), which indicates a significantly weak relationship between these two parameters. The decay index of AR and QS filaments are very close to each other. Their average values are $\sim$0.8 and $\sim$0.9, respectively. More interestingly, we find that the decay indices have an upper limit of $\sim$1.4 and an average value of $\sim$0.9 for all events, which are very close to the critical decay index (1.1--1.3) for the torus instability of a deformable and thick MFR as expected in the solar corona \citep{Demoulin2010Criteria}. It indicates that these filaments at the near-eruption stage are approaching the critical height, where the torus instability may take place to initiate their final fast eruption phase.

\section{Summary and Discussions}
In this Letter, we perform a statistical study on the long-duration quasi-static evolution of filaments, mainly focusing on the evolution of their heights in 3D. We reach a new finding that the filaments exhibit two different types of long-duration quasi-static evolution. The evolution for the group 1 filaments is comprised of a long-duration static phase and a short-duration slow rise phase, while for the group 2 filaments, it only appears a slow rise phase but with an extremely long duration of at least 3 days. We also study the physical parameters of all filaments at the near-eruption stage including the real height, length, background magnetic field strength and decay index. It is found that the filament height is linearly related with the filament length and the background magnetic field strength is exponentially related with the filament height. However, the decay index is independent of any filament parameters and distributed in a narrow range with an upper limit of $\sim$1.4 and an average value of $\sim$0.9. 

The fact that the pre-eruptive filaments present two different types of long-duration quasi-static evolution indicates there may exist two distinct physical mechanisms to control their quasi-static evolution. The group 1 filament events are similar to the events previously studied by \cite{2004ApJ...602.1024S} and \cite{McCauley2015Prominence}. A common characteristic is that the filaments experience a slow rise with a duration of less than 30 h. The average speed during the slow rise phase is usually several km s$^{-1}$. Such a slow rise is usually thought to be due to breakout reconnection occurring above the filaments or tether-cutting reconnection below \citep{2004ApJ...602.1024S,2005ApJ...630.1148S}, even sunspot rotation \citep{2013SoPh..286..453T} and flux transmission \citep{2012ApJ...756...59L}. However, for the group 2 filaments, the slow rise phase (at least 3 days) is much longer than the events previously studied \citep[0--14 h;][]{McCauley2015Prominence}. As expected, the long duration of the slow rise phase corresponds to a very small speed (0.01--0.2 km s$^{-1}$). A completely new mechanism that is able to interpret the extremely long slow rise phase and very small speed should be considered in the future.

The parameters of the filaments (height, length, and background magnetic field strength) are continuously distributed if the AR and QS filaments are not differentiated, which indicates that AR and QS filaments may have a similar magnetic structure. Prior to the eruption, both AR and QS filaments even have a similar average decay index. The upper limit and average value of the decay index are $\sim$1.4 and $\sim$0.9, respectively. They are comparable with the theoretical threshold (1.5) for torus instability of a freely expanding MFR \citep{2006PhRvL..96y5002K} and that (0.5--2.0) of partial torus instability for a partial MFR with two footpoints anchored in the photosphere \citep{2010ApJ...718..433O}. They are even closer to the critical decay index of 1.1--1.3 for a deformable and thick MFR as expected in the solar corona \citep{Demoulin2010Criteria}. The similarity of the observational and theoretical decay indices suggests that the magnetic structure could be a magnetic flux rope, whose fast eruption is initiated by the torus instability. 

It is worthy of noticing that the PFSS model we used has its limitations. First, the bottom boundary is based on Carrington synoptic maps, which ignores the variation of the magnetic field in the filament source regions. Secondly, the assumption of the PFSS model that the background magnetic field is potential one may be oversimple because the other current system besides the filament may also has an important contribution to the background field \citep{2011ApJ...732...87C}. However, for the present statistical study, in particular when calculating the large-scale magnetic field over the QS filaments, the PFSS model is an optimal choice. It is noticed that \cite{Liu2008Magnetic} also used the PFSS model and obtained an average critical decay index of $\sim$1.7 in the height range of 42-105 Mm. Here, we derive a relatively smaller critical decay index with a more accurate estimate of the filament height. It is also worth noticing that the magnetic structure of filaments could be sheared arcades \citep{2005ApJ...629.1122D}. However, according to a recent statistical study for 571 erupting filaments by \cite{2017ApJ...835...94O}, it is found that 89\% filaments are an MFR configuration, whereas only 11\% cases show a sheared arcade structure. The possibility of few cases with a sheared arcade structure thus does not influence our main conclusions.

\acknowledgements We are grateful to the referee for his/her constructive comments that significantly improved our manuscript. AIA data are courtesy of NASA/SDO, a mission of NASA's Living With a Star Program. This work is funded by NSFC grants 11722325, 11733003, 11790303, Jiangsu NSF grants BK20170011, and ``Dengfeng B" program of Nanjing University.

\end{document}